# Energy Efficient Service Placement for IoT Networks

**Mohammed A. Alshahrani, Ahmad Adnan Qidan, Taisir E. H. El-Gorashi, and Jaafar M. H. Elmirghani**

*Department of Engineering, Faculty of Natural, Mathematical & Engineering Sciences, King's College London*
E-mail:{mohammed.alshahrani@kcl.ac.uk, ahmad.qidan@kcl.ac.uk, taisir.elgorashi@kcl.ac.uk, jaafar.elmirghani@kcl.ac.uk}

**ABSTRACT**
In recent years, there has been a significant expansion in the Internet of Things (IoT), with a growing number of devices being connected to the internet. This has led to an increase in data collection and analysis as well as the development of new technologies and applications. The rise of IoT has also brought about new challenges, such as security concerns and energy efficiency. This study investigates a layered IoT architecture that combines fog and cloud computing, aiming to assess the impact of service placement on energy efficiency. Through simulations, we analyse energy use across Access Fog, Metro Fog, and Cloud Data Centre layers for different IoT request volumes. Findings indicate that Access Fog is optimal for single requests, while Metro Fog efficiently manages higher demands from multiple devices. The study emphasizes the need for adaptive service deployment, responsive to network load variations, to improve energy efficiency. Hence, we propose the implementation of dynamic service placement strategies within Internet of Things (IoT) environments.
**Keywords**: IoT, Fog, Cloud, Service Placement, Energy Efficiency.

## 1. INTRODUCTION

The ubiquity, intelligence, actuating capabilities, and wireless connectivity of the Internet of Things (IoT) are used as the foundation for designing and creating smart domains, such as smart cities, smart buildings, and smart transportation. With the number of IoT devices estimated to reach 75.44 billion globally by 2025, there has been a corresponding exponential increase in the demand for data communication and processing infrastructures [1]. This surge necessitates a fundamental shift in how data is managed, and services are deployed, significantly increasing data processing transferred to the cloud. However, this paradigm shift incurs considerable energy requirements due to the extensive transmission of data and the computational workload on cloud servers [2]. Managing the energy required to sustain the connections of billions of devices and the processing and storage of vast amounts of data represents a significant challenge. This highlights the urgent need for energy-efficient service placement strategies capable of mitigating the environmental impact and operational costs of IoT systems [3]. The advent of edge computing has offered a promising approach by processing data closer to its source. Nonetheless, realizing the goal of optimal energy efficiency in service placement continues to be a critical challenge for researchers [4].

Fog computing, which extends cloud computing, strategically situates processing power near data origins via edge devices. This approach not only enhances cloud functionality but also complements it, providing a robust framework for developing advanced IoT applications and services [3]. By integrating fog computing into IoT systems, we can address the scalability and sustainability issues associated with relying solely on cloud computing. This method allows for the simultaneous optimization of the entire network infrastructure, including end devices and core computing resources, thereby significantly improving performance and efficiency [5]. Our previous research on optimizing energy efficiency in cloud and core networks examined diverse scenarios, such as integrating renewable energy, optimizing network topology, managing data centre networks, enhancing content delivery, applying network coding, and implementing virtualization.[6]-[12].

In this paper, we examined the impact of service placement on energy efficiency within IoT architectures, analysing how different computational layers Access Fog, Metro Fog, and Cloud DC respond to varying IoT device requests. Through detailed simulations, we identified the optimal layers for processing single and multiple requests, highlighting the importance of adaptive service strategies for energy conservation.

The remainder of this paper is organized as follows: section 2 discusses energy-efficient data processing methods across cloud, edge, and fog computing. Section 3 outlines our proposed multi-layer IoT architecture. The results and discussions are demonstrated in section 4. The conclusions are presented in Section 5.

## 2. Energy-Efficient Data Processing in IoT

Effective data processing methods are crucial for minimising energy consumption, which in turn prolongs the lifespan of IoT devices and reduces their environmental footprint. Various architectural paradigms, including cloud computing, edge computing, and fog computing, have emerged to address the demand for efficient IoT data processing, each with its unique benefits and limitations. Cloud Computing has traditionally served as the primary support for processing IoT data, providing strong computational capabilities and significant storage space. The centralised structure makes management easier and enables advanced data analysis. However, centralization presents issues such as excessive latency caused by the geographical separation of IoT devices and cloud servers, as well as substantial energy consumption for data transfer. These problems highlight the need for alternate paradigms that can reduce these shortcomings [13]. Edge Computing distributes data processing, which decreases the requirement for data to be transferred to and from the cloud. The closeness to data sources significantly reduces latency and decreases the energy needed for data transmission, offering an effective option for real-time applications and improving data privacy. However, this method has limits, especially regarding the processing and storage capacities of edge devices in comparison to centralised cloud data centres [4].

   Fog Computing acts as a mediator that combines the cloud's computing power with the edge's closeness to data sources. Fog computing provides a balanced approach by placing processing nodes closer to IoT devices than edge computing, but not as spread. It utilises localised processing to decrease latency and save bandwidth, while still allowing for intricate processing jobs across a distributed network of fog nodes. The combination of such features makes fog computing an attractive choice for enhancing energy efficiency in IoT devices [3]. Overall, the decision on the architectural design for data processing in IoT is impacted by a variety of aspects such as energy efficiency, latency, processing power, and scalability. The interaction of cloud, edge, and fog computing in IoT networks will influence energy-efficient data processing developments to meet the growing demands of a hyper connected world.

## 3. System Model

The network discussed in this study consists of three layers: the access network, the metro network, and the core network. The structure encompasses four main layers, illustrated in Figure 1, forming a comprehensive network that connects IoT devices directly to the central cloud. The initial layer is made up of IoT devices, which are smart wireless nodes used to collect and forward data to upper layers for further processing and storage. The Access Fog layer consists of wireless access points that collect service requests from the IoT devices. These requests are either transmitted to the next layer(s) or processed locally. The metro network comprises a high-capacity Ethernet switch and a pair of edge routers, serving as a gateway to cloud data centres via the core network. [14]. Positioned on the right, the Cloud layer houses data centres equipped with high-capacity servers.

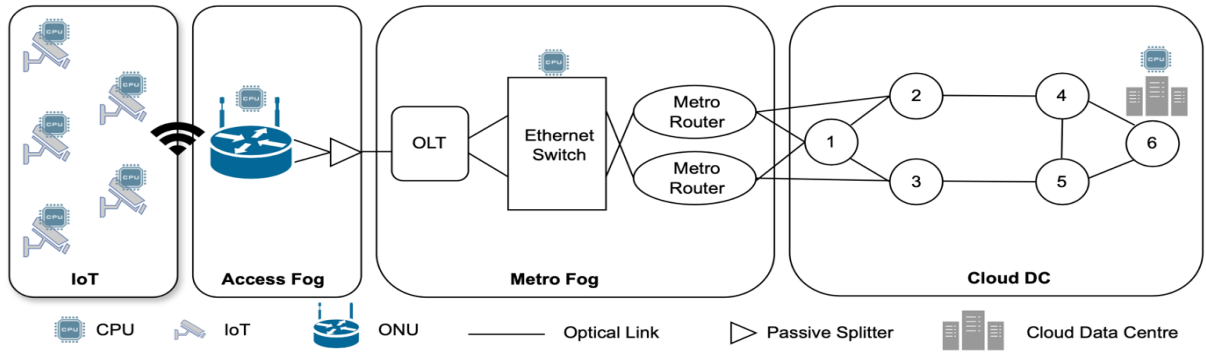

*Figure 1. IoT-Fog-Cloud layers architecture*

## 4. Results and Discussions

In this paper, we consider IoT devices generating service requests, and we assume that there is no task splitting between the computing layers. Each demand consists of processing request in (MIPS) and data rate in (Mbps). We evaluate the efficiency of distributed processing models for energy conservation using a setup where 5 IoT devices are linked to the access network via a single Optical Network Unit (ONU). This setup imitates a typical home LAN network, which generally supports a limited number of users [15]. The processing requirements data was obtained from the study conducted in [16], which states that a standard analysis and decision-making application in the IoT necessitates 750 instructions for the processing of a single bit of data. By employing a direct methodology, we estimate that the processing of 1MB of traffic requires around 1000 MIPS. The allocation of 1000 MIPS is designated for the processing of each unit of traffic, with the capability to handle video resolutions ranging from 1024 x 720 to 1600 x 1200, while maintaining a frame rate of 30 frames per second. [17]. Table 1 and 2 list All the parameters of the networking devices and processing servers set according to other studies in [21]-[26].

*Table 1. Input Parameters for network components.*

| Device | Bitrate(Gbps) | Pmax (W) | Pidle (W) | Device Layer |
|---|---|---|---|---|
| IoT (Wi-Fi) | 0.15 | 0.56 | 0.34 | IoT |
| ONU (Wi-Fi) | 0.3 | 15 | 9 | Access Fog |
| Metro Router Port | 40 | 30 | 27 | Metro Fog |
| Metro Ethernet Switch | 600 | 470 | 423 | Metro Fog |
| IP/WDM | 40 | 1.15k | 1k | Core Network |

*Table 2. Input Parameters for processing servers.*

| Device | Pmax (W) | Pidle (W) | GHz | MIPS | Device Layer |
|---|---|---|---|---|---|
| RPi Zero W | 3.96 | 0.5 | 1 | 1000 | IoT |
| RPi 3 | 12.5 | 2 | 1.2 | 2400 | Access Fog |
| Intel X5675 | 95 | 57 | 3.06 | 73440 | Metro Fog |
| Intel Xeon E5-2680 | 130 | 78 | 2.7 | 108000 | Cloud |

The CPU capacity per Access Fog device was assessed using a methodology similar to the IoT, measured in MIPS. In order to accommodate the varying efficiency provided by the two layers, we used the assumption that Access Fog devices have a greater CPU capacity compared to IoT devices. The Ethernet switch is typically situated in facilities that facilitate connectivity to public cloud services. Its primary function is to gather traffic from one or many access networks. Typically, metro routers are responsible for the control of traffic and authentication [18]. The computational capabilities of the metro fog exceed those of the lower fog layers, enabling it to accommodate more users and services, yet they do not match the capacity of the cloud DC [19]. The core network operates as an IP/WDM network with two distinct layers: the IP layer, where routers are strategically placed at each node to streamline and oversee network traffic, and the optical layer, which is designed to connect IP routers through optical switches and IP/WDM technologies [20]. It is assumed that a centralised cloud data centre is situated at node 6 in Fig. 1, with an unrestricted quantity of compute servers. Networking and processing equipment energy usage profile consists of two components: an idle component and a load-dependent component. The energy consumption related to load-dependent activities can be determined by employing the metrics of energy per bit for communication and energy per instruction for processing.

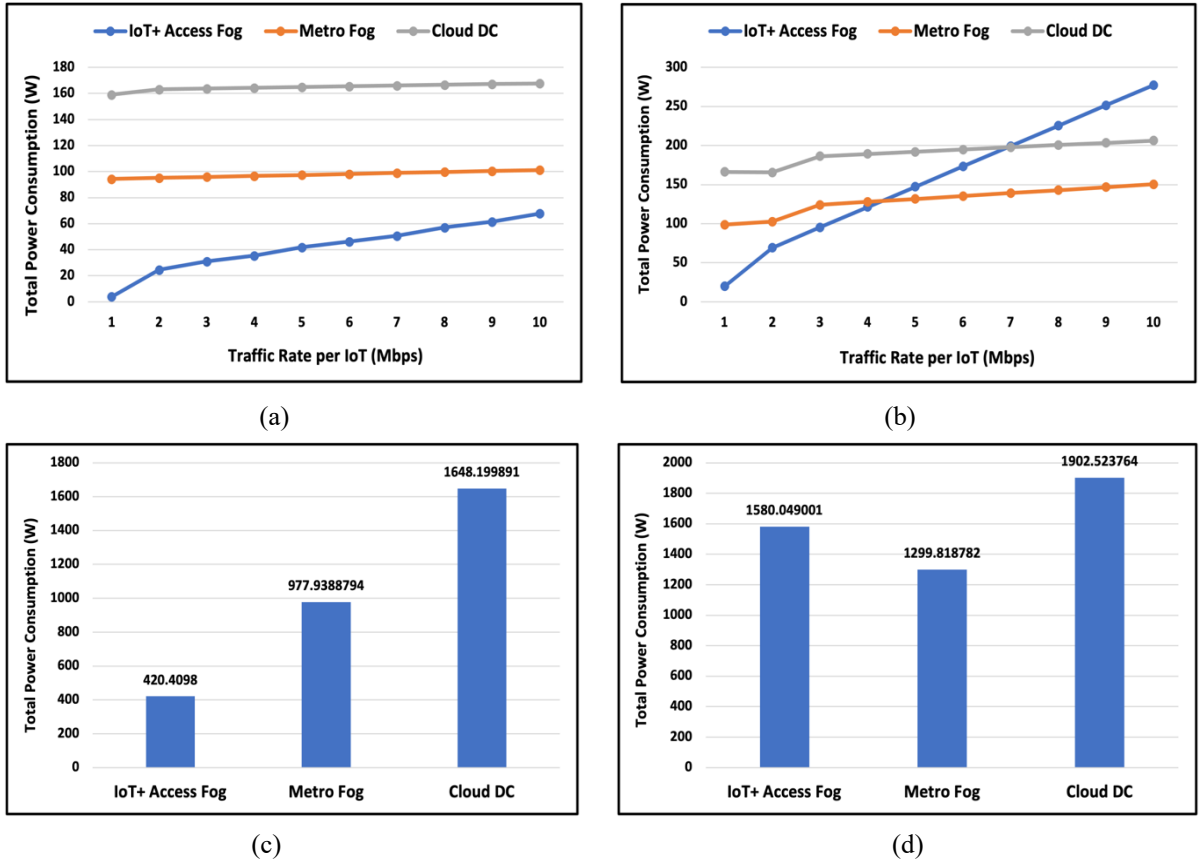

*Figure 2. (a) Power consumption for scenario 1 (b) Power consumption for scenario 2 (c) Comparison of total power consumption for scenario 1 (d) Comparison of total power consumption for scenario 2*

This study helps understanding the implications of service placement strategies on the system's energy efficiency under two different load scenarios. In the first scenario, the total power consumption for processing a single IoT request is evaluated across the Access Fog, Metro Fog, and Cloud DC layers as shown in Figure 2 (a). It is shown that the power consumption of the Access Fog layer demonstrates a gradual rise as the traffic rate per IoT device increases. This is expected, as more data processing at the edge results in a proportional energy cost due to the limited computational efficiency of edge devices. However, the power consumption remains the lowest across all layers. The Metro Fog layer maintains a steady, moderate power consumption level across all traffic rates. Its flat energy consumption profile indicates that Metro Fog nodes are designed to handle varying traffic rates without significant fluctuations in energy use, reflecting their suitability for moderate load scenarios. Cloud DC shows the highest but consistent energy consumption across different traffic rates. The substantial computational resources of the Cloud DC, while power-hungry, are not significantly affected by the traffic rate changes in this scenario. This implies that cloud resources are under-utilized for single IoT request processing, making it an energy-inefficient choice for such low-scale tasks.

The second scenario, reflected in Figure (b), presents the energy consumption for processing requests from five IoT devices. Here, the trend is markedly different. The Access Fog layer shows a substantial increase in energy consumption, although it remains the lowest among the layers. This indicates that while edge processing remains more energy-efficient than cloud processing, its relative efficiency decreases as the load increases. The Metro Fog layer's power consumption rises significantly with the increase of traffic load, which shows that its performance is compromised under heavier loads. In contrast, the Cloud DC layer exhibits the highest energy consumption, which increases steeply with the higher traffic rate per IoT, due to the fact that the cloud's scalable resources handle increased loads at the cost of high energy consumption.

According to Figure 2 (c), the Access Fog layer is the most energy-efficient for processing a single IoT request, while energy consumption increases at the Metro Fog and Cloud DC layers if the processing is occurred at any of them. Figure 2 (d) demonstrates that although Access Fog remains the least energy-intensive option for five IoT queries, its power consumption increases in a non-linear manner. Metro Fog has more effective scalability, while Cloud DC exhibits significantly higher power usage, indicating a comparatively lower level of energy efficiency when handling many requests.

## 5. CONCLUSIONS

In this paper, we demonstrated that energy efficiency in IoT service placement varies significantly across computational layers, with the Access Fog layer enhancing performance for single requests, and the Metro Fog layer showing greater efficiency for handling the demands of five IoT devices. The results indicate a critical need for strategic service placement in IoT architectures to efficiently control energy consumption in different network loads.

## ACKNOWLEDGEMENTS

This work has been supported by the Engineering and Physical Sciences Research Council (EPSRC), in part by the INTERNET project under Grant EP/H040536/1, and in part by the STAR project under Grant EP/K016873/1 and in part by the TOWS project under Grant EP/S016570/1. All data are provided in full in the results section of this paper. The first author would like to acknowledge the Government of Saudi Arabia and King Khalid University for funding his PhD scholarship.